%% file: main.tex
\documentclass{vldb}

\usepackage{balance}  %
\usepackage{booktabs}
\usepackage{url}
\usepackage{dblfloatfix}
\usepackage{float}
\usepackage{makecell}

\graphicspath{{./figures/}}
\usepackage{graphicx}
\usepackage[font=bf,labelfont=bf]{caption}
\usepackage{subcaption}
\usepackage{tikz}
\usetikzlibrary{positioning,shadows,shapes,arrows}
\usetikzlibrary{arrows.meta,chains,matrix,decorations.pathreplacing}
\usepackage{pgfplots}
\usetikzlibrary{arrows.meta,
                calc, chains,
                positioning,
                shapes.multipart}
\usepackage{listings}

\usepackage{caption}

\newlength\myindent
\usetikzlibrary{matrix,positioning,arrows.meta,arrows}
\usepackage{pdfpages}

\pgfplotsset{compat=1.16}
\usetikzlibrary{decorations.pathreplacing,calligraphy}

\usepackage{algorithm}
\usepackage{algorithmic}

\usepackage{xspace}
\newcommand*{\eg}{e.g.,\@\xspace}
\newcommand*{\ie}{i.e.,\@\xspace}

\usepackage{xcolor}

\newcommand{\sparagraph}[1]{\vspace{1mm}\noindent {\bf #1}}

\vldbTitle{Bounding the Last Mile: Efficient Learned String Indexing}
\vldbAuthors{Benjamin Spector, Andreas Kipf, Kapil Vaidya, Chi Wang, Umar Farooq Minhas, Tim Kraska}

\begin{document}

\title{Bounding the Last Mile: Efficient Learned String Indexing (Extended Abstracts)}

\numberofauthors{6} %

\author{
\alignauthor
Benjamin Spector\\
\affaddr{MIT CSAIL}\\
\email{spectorb@mit.edu}

\alignauthor
Andreas Kipf\\
\affaddr{MIT CSAIL}\\
\email{kipf@mit.edu}

\alignauthor
Kapil Vaidya\\
\affaddr{MIT CSAIL}\\
\email{kapilv@mit.edu}

\and

\alignauthor
Chi Wang\\
\affaddr{Microsoft Research}\\
\email{wang.chi@microsoft.com}

\alignauthor
Umar Farooq Minhas\\
\affaddr{Microsoft Research}\\
\email{ufminhas@microsoft.com}

\alignauthor
Tim Kraska\\
\affaddr{MIT CSAIL}\\
\email{kraska@mit.edu}

}

\maketitle

\input{abstract}
\input{introduction}

\input{approach}
\input{evaluation}
\input{conclusions}

{
\setlength{\parskip}{0.5em}
\footnotesize
\sparagraph{Acknowledgments.}
This research is supported by Google, Intel, and Microsoft as part of DSAIL at MIT, and NSF IIS 1900933. This research was also sponsored by the United States Air Force Research Laboratory and the United States Air Force Artificial Intelligence Accelerator and was accomplished under Cooperative Agreement Number FA8750-19-2-1000. The views and conclusions contained in this document are those of the authors and should not be interpreted as representing the official policies, either expressed or implied, of the United States Air Force or the U.S. Government. The U.S. Government is authorized to reproduce and distribute reprints for Government purposes notwithstanding any copyright notation herein.
\par
}

\balance

\bibliographystyle{abbrv}
\bibliography{main}

\end{document}

%% file: abstract.tex
\begin{abstract}

We introduce the RadixStringSpline (RSS) learned index structure for efficiently indexing strings. RSS is a tree of radix splines each indexing a fixed number of bytes. RSS approaches or exceeds the performance of traditional string indexes while using 7-70$\times$ less memory. RSS achieves this by using the minimal string prefix to sufficiently distinguish the data unlike most learned approaches which index the entire string. Additionally, the bounded-error nature of RSS accelerates the last mile search and also enables a memory-efficient hash-table lookup accelerator. We benchmark RSS on several real-world string datasets against ART and HOT. Our experiments suggest this line of research may be promising for future memory-intensive database applications.

\end{abstract}

%% file: introduction.tex
\section{Introduction}

Learned indexing as introduced by \cite{learnedindexes} has brought new perspective to indexing by reframing it as a cumulative distribution function (CDF) modeling problem. The burgeoning field, despite its nascence, has brought with it many opportunities and efficiencies. However, most work in this area has focused on efficiently indexing numerical keys. There are several reasons why building learned indexes for strings is usually harder than for numerical keys:

\begin{itemize}
    \item Strings are variable-length, complicating both model inference and memory layout.
    \item Strings are much larger objects which are expensive to store, compare, and manipulate.
    \item Strings tend to have even worse distributional properties than integer keys due to the prevalence of common prefixes and substrings, making them exceedingly hard to model accurately and compactly.
\end{itemize}

We also distinguish between two indexing cases: (1) primary index: the data is sorted according to the key and the index is used to more efficiently find the data inside the sorted data, and (2) secondary index: the data itself is not sorted and the index has to store pairs of keys and tuple identifiers (TIDs) in the leaf nodes. 
Commonly-used string-key secondary index structures include B-trees and their variants \cite{btree}, ART \cite{art}, HOT \cite{hot}, and the highly space-efficient FST \cite{surf}; learned indexes are absent. Recent learned primary index structures include SIndex~\cite{sindex} (also operating on strings), ALEX~\cite{alex} and Hist-Tree~\cite{cht}.

In our view, the most significant problem with learned indexes for strings is the last-mile search which corrects the learned model's prediction. Last-mile search is usually implemented as an exponential search around the prediction which expands out until a bound is found, followed by a binary search inwards to find the true index. This search is especially expensive in string scenarios for two reasons. First, average model error in these scenarios is often quite high, due to difficulty in modeling real-world datasets. Because many real-world datasets have both long shared prefixes and relatively low discriminative content per byte, the CDF appears to be almost step-wise from afar, which is difficult for traditional learned models to accurately predict or capture.
Second, actually conducting the last mile search is slow. Each comparison is expensive and requires potentially many sequential operations, and the strings’ large size decreases the number of keys which fit in cache.
Modern column stores typically store strings in fixed length sections (first few bytes) and variable length sections (remainder of the string) \cite{umbra}. Increasing the size of the fixed-length section can waste memory, and searching the variable length section will incur expensive cache misses.
One clear improvement, which forms the basis of this research, comes from having bounded error (\ie the model outputs a bounded interval in which the sought key will lie, if present). Then, one can replace the exponential search with binary search and skip many string comparisons. %
We use this to ameliorate the cost of the last-mile search for lower bound lookups (\ie finding the key that is equal to or larger than the lookup key), and then, for equality lookups, we bypass it (more later).

In this work, we emphasize increasing the lookup performance and decreasing the size for the primary-index scenario.
For example, such an index could be used for global dictionary encoding \cite{carstenpaper}.
We introduce the RadixStringSpline (RSS), a learned string index consisting of a tree of the learned index structure RadixSpline (RS) \cite{radixspline, sosd-neurips, sosd-vldb}.
RS is a learned index that consists of an error-bounded spline which is in turn indexed in a radix lookup table.
Similar to nodes in ART, each RSS node indexes a fixed partial key (\eg 8 bytes). But unlike ART, our nodes may have extremely high fan-out due to the incorporation of a learned-model at each node.
To increase the fan-out of a given byte-prefix, we encode the input data using HOPE \cite{hope}.
HOPE eliminates common prefixes and increases information density.
Like the original string data, HOPE-encoded data has variable length.
For a popular URL dataset, HOPE encoded-data is on average 1.6$\times$ smaller than the original string data.
Often, the first 8 bytes of the HOPE encoding are sufficient to distinguish between most of the string keys.
RSS only requires a single node to index these keys.
However, typically there also are many collisions among the 8-byte prefixes, which requires RSS to recurse on the following 8 bytes.

While we do not discuss updates, techniques as proposed in ALEX~\cite{alex} are generally applicable to our approach.

Compared to ART and HOT, RSS is faster to build, similarly fast to query, and consumes much less memory (7-70$\times$). %
RSS particularly benefits from a high information density in the most significant bytes. Some data distributions, like the Twitter Semantic140 dataset, satisfy this well, and correspondingly RSS has high performance. For other data distributions (\eg URLs) that require many low-information bytes to distinguish keys, RSS needs many levels and hence may have low performance.

%% file: approach.tex
\section{Splining Strings}
\label{sec:triespline}

\sparagraph{Problem Statement.} Given an immutable, lexicographically sorted array of strings, we want to build an index on top which supports two key operations. First, the index should support equality lookups: if the string is present, return its index, and if not, return NULL. Second, the index must support lower bound queries: given a string, find the index of the greatest string greater than or equal to the provided one.
These two operations are important in column stores that use dictionary encoding.
For equality queries (\ie WHERE str = X) one needs to support equality lookups on the dictionary, and for prefix queries (\ie WHERE str LIKE 'A\%') lower bound lookups are required (to find the first string in the dictionary that is lexicographically greater than or equal to 'A').

\sparagraph{Method.} Our method consists of two parts. First, we describe the RadixStringSpline (RSS), a compact and efficient adaptation of RadixSpline to the string domain \cite{radixspline}. This approach provides both fast queries as well as bounded error. Second, we provide an optional add-on hash corrector which makes use of this bounded error to, at the cost of 12 bits per element, further improve equality lookup performance.

\sparagraph{RadixStringSpline.} One useful perspective of the order-preserving indexing problem is that of a compressive mapping. If a million keys use a 64 bit integer type, the overall density of information is quite low relative to the capacity of the type. Indexing transforms these million keys into a continuous range, effectively compacting the data into the last 20 bits.

\begin{figure}[H]
    \centering
    \includegraphics[width=.75\linewidth]{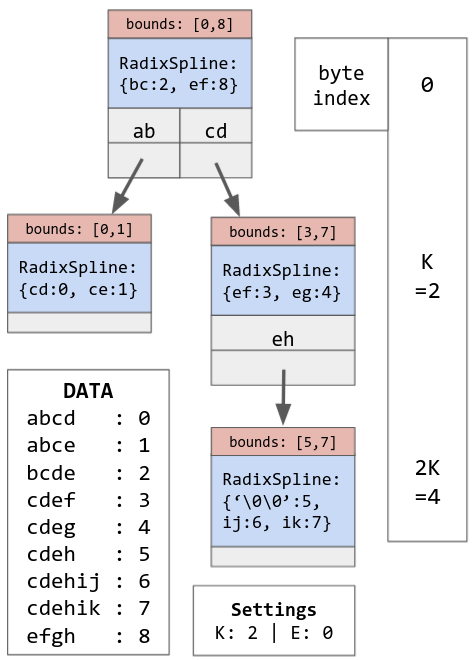}
    \caption{A sample RSS tree structure, for the toy data and settings indicated. The root node corresponds to the entire data (bounds [0,8]) and indexes the first two bytes of the data (K=2). The RadixSpline in the root indexes the only two keys which do not have collisions in the first two bytes (bc and ef). The remaining two-byte prefixes (ab and cd) have collisions and are redirected to child nodes.}
    \label{fig:rssdiagram}
\end{figure}

The core difference between string indexing and integer indexing is that integer keys have much higher entropy in the first few bytes than strings do because integer keys must be entirely distinct in this range. By contrast, string keys can (and in most practical applications frequently) share long prefixes, requiring examination of more data to fully distinguish them. There are several (not necessarily exclusive) approaches one might take to ameliorate this. For example, one can actually directly compact the data, as per \cite{hope}, and this certainly does help. In our approach, our goal is to have the model quickly operate on the minimum amount of data to get the job done.

An RSS is a tree, in which each node contains:
\begin{itemize}
    \item The bounds of the range over which it operates.
    \item A redirector map. This contains the keys for which the current node cannot satisfy the error bound, and pointers to context-aware nodes for each of those keys.
    \item A RadixSpline model operating on K bytes, with an error-bound of $E$.
\end{itemize}

We illustrate a sample RSS in Figure \ref{fig:rssdiagram}. The RSS indexes the indicated data in the bottom left from 0 to 8, with each node operating on two-byte chunks and a maximum allowable error of 0. (We note that practical RSS's usually have a greater allowable error but these are harder to intuitively illustrate.) All searches begin at the root node (top) and simultaneously traverse the tree and the string until the partial key can no longer be found in the redirector array. At that point, the local RadixSpline at that node is guaranteed to provide a bounded-error prediction.

To build an RSS, one begins by building a RadixSpline on the first K bytes of every string in the dataset (this is the root node), including all duplicates. Then, we iterate through the unique K-byte prefixes, and check if the estimated position is within the prescribed error bounds for both the first and last appearance of the prefix. This will always be the case for unique prefixes but might not be for prefixes which have duplicates, and cannot be the case for prefixes which have $>2E$ duplicates -- then, even predicting the median instance can't satisfy the extrema. For each prefix which fails the test, we add it to the redirector table, and build a new RSS over just the range of the problematic prefix and starting at byte K instead of byte 0. This process continues recursively until every key is satisfied.

We note that the radix table in the RadixSpline should be adjusted depending on where in the tree one is. Near the root, the radix table should be large; near the leaves we often use just 6 bits to save memory.

Practically we have found K=8 or K=16 and E=127 to be good settings; in our experiments we use K=16 (via gcc's builtin \_\_uint128\_t type) since it is more robust to sparser data. For simplicity we keep these parameters fixed; pragmatically there are probably reasonable improvements to be had in both memory and query time by switching to K=8 for smaller, easier-to-model datasets.
Note that a fanout spanning 16 bytes is generally far higher than the maximum fanout of ART (1 byte, \ie 256 keys, per level) or HOT (32 keys per level).

Querying an RSS is simple. One begins by extracting the first K bytes of the string, and conducting a binary search of the redirector to try to find the prefix. If it is found, then one follows the redirect to the new RSS node and the process begins again, only operating on the next K bytes. If it is not found in the redirector, then that means the key is guaranteed to be within acceptable bounds for the RadixSpline, so one queries the RadixSpline at the current node with the appropriate substring and returns the result.

As an example, suppose we want to query the string ``cdeg'' from the RSS in Figure 1. We begin by extracting the first two bytes of the string, in this case ``cd'', which are packed into a 16-bit integer. We then search the redirector of the root node for this key. Since we find it in the second slot, we follow the pointer at that slot to the next node of the RSS. Now, we extract the next two bytes, ``eg'' and check the node's redirector. This time, we fail to find it, so we know it is correctly indexed by the local RadixSpline. So, we query the RadixSpline and return the result as our prediction. Finally, we execute a local binary search in the data to either find the string and its index or else to validate its non-existence.

Analogously, suppose we wanted to conduct a lower bound query for the string ``defg'', not found in the data. We again search the root node's redirector for ``de'', and, failing to find it, execute the root's RadixSpline on that prefix. We again perform a binary search, only this time after failing to find it we return the left bound.

One additional benefit of our approach is that if it is appropriately constructed (requiring a synchronization of the predictions of different layers of the tree) it can also be made perfectly monotonic, which may prove useful in future work on accelerating the last mile search.

We believe that one should not undervalue the importance of the model being error-bounded for two reasons. First, in a string setting, even with relatively low errors and an optimized last-mile search, the last mile search still turns out to be the dominant cost. A bounded error means one only needs to conduct a binary search rather than an exponential search. Second, it enables a memory-efficient hash corrector, to be described below.

\sparagraph{Hash Corrector.} Often, while index structures certainly need to support lower bound queries as described above, the optimization of the actual direct lookup of known string keys is equally important. To this end, we provide an auxiliary data structure which can improve performance for this problem at the cost of a small amount of additional memory. Essentially, one stores a contiguous array of signed int8 offsets, with -128 reserved as empty. To build the HC, for each string in the dataset one runs the RSS and computes the difference between the predicted and true values, which is guaranteed to fit in the range -127 to 127. Then, one hashes the string into these slots several times (up to a predetermined number) to find an empty slot. If one is found, we insert the offset at that slot.
Compared to traditional Cuckoo hashing, this technique trades false positives (\ie the string at the offset does not match the lookup key) for memory efficiency.
When one queries a string, one again hashes the string to a few of these slots and tries each offset. If it's a match, the expensive binary search is avoided. Otherwise, the bounded binary search is a reasonable fall-back (this is also required for negative lookups, although we deem them to be less important in practice (\eg in the dictionary encoding scenario).
To have some benefit from false positive lookups, our implementation uses the keys it finds at these offsets to at least reduce the bounds of the binary search, and these reduced bounds can also help us rapidly reject wrong offsets (which are out of the current bounds). So, each query to the underlying data is guaranteed to provide at least some benefit.

In our implementation, we use a 128-bit MurmurHash3 hash \cite{murmurhash3}, \cite{murmurhash3code} giving us 4 attempts, and we set the load factor to be 2/3. This then provides a speed boost to >95\% of lookup queries at the cost of 12 bits per key. Further tuning this time-space trade-off might lead to additional gains.

For example, suppose we want to look up a string S in a database of N strings. We first execute the RSS to get an error-bounded index prediction $p$. We then hash S into 4 positions in range [0, 3N/2) -- $h_1$, $h_2$, $h_3$, $h_4$. For each position $h_i$, if offsets[$h_i$] == -128 or exceeds the bounds, we immediately know it is invalid and skip it. Otherwise, we compare S to the string at $p$+offsets[$h_i$]. If they match, we return; if S is larger, we set the location as a left bound, and if S is smaller we set the location as a right bound. Finally, if we try four times and still have not found it, we execute a binary search between the left and right bounds. The data structure can and should be entirely ignored for lower bound queries; it does not currently accelerate them.

\pagebreak

%% file: evaluation.tex
\section{Evaluation}
\label{sec:evaluation}

We evaluate RSS against ART and HOT as baselines on four datasets:
\begin{itemize}
    \item Wiki: $>$13M unique Wikipedia URL tails. \cite{wiki}
    \item TwitterSentiment: 1.6M tweets, meant to provide representative natural language. \cite{twittersentiment}
    \item Examiner: 3M headlines from the Examiner. \cite{examiner}
    \item URL: Approximately 100M URLs from a 2007 web crawl; approximately 10GB total. \cite{url}
\end{itemize}

\begin{table}[]
\centering
\resizebox{\linewidth}{!}{
\begin{tabular}{l|l|l|l|l}
\textbf{Build, ns/item} & Wiki & Twitter & Examiner & URL \\
\hline
ART    & 131  & 147     & 143      & 197 \\
\hline
HOT    & 207  & 209     & 217      & 243 \\
\hline
RSS    & \textbf{42}   & \textbf{41}      & \textbf{37}       & \textbf{94}  \\
\hline
RSS+HC & 171  & 200     & 184      & 383 \\
\hline
\hline
\makecell{\textbf{Lookup, ns} \\ \textbf{(LowerBound)}}  & Wiki      & Twitter   & Examiner  & URL       \\
\hline
ART    & 785       & 530       & 666       & 1592      \\
\hline
HOT    & 494 (\textbf{584}) & \textbf{365} (472) & \textbf{394} (\textbf{514}) & \textbf{786} (\textbf{920}) \\
\hline
RSS    & 629       & 452       & 554       & 1733      \\
\hline
RSS+HC & \textbf{477} (629) & 378 (\textbf{452}) & 427 (554) & 1314 (1733) \\
\hline
\hline
\textbf{Memory, MB} & Wiki   & Twitter & Examiner & URL      \\
\hline
ART    & 1219.8 & 223.5   & 374.7    & 15,573.0 \\
\hline
HOT    & 205.9  & 24.7    & 48.3     & 1,372.3  \\
\hline
RSS    & \textbf{3.6}    & \textbf{1.1}     & \textbf{1.6}      & \textbf{198.8}    \\
\hline
RSS+HC & 23.8   & 3.4     & 6.1      & 350.7     
\end{tabular}}
\caption{Results for ART, HOT, and the RadixStringSpline with and without its add-on Hash Corrector. Note that for the second sub-table, parenthetical values are for lower bound queries if they are appreciably different from lookup queries.}
\label{tab:mainresult}
\end{table}

We summarize the results in Table~\ref{tab:mainresult}. The RSS is generally faster than ART but slower than HOT, but far smaller (7-70$\times$) than either of them. With its hash corrector, lookup speed is usually comparable to HOT and the data structure remains smaller, although memory usage increases somewhat. The RSS is also extremely fast to construct compared to existing structures, with a speed boost of 2-3$\times$, although this is at the expense of not supporting inserts.
However, the fast construction time emphasizes that RSS is particularly useful for bulk-loading and delta-updates.
We take note of RSS's poor performance on the URL dataset; we discuss this below.

The performance character of RSS is distinguished from traditional trie-like data structures in that its compound nodes have unlimited fanout; what decides the cost of the model is the depth of the data required, since each inner node requires another local redirector query to find a new node with the additional context. To this end, the URL dataset is practically an adversarial scenario (as would be a filesystem) -- virtually all strings share a relatively small number of long prefixes, which lowers the discriminatory capacity of the local spline. To validate this understanding, we run the same datasets, only this time encoded by HOPE's two-gram approach in order to localize more data at the start of the model. The result is a considerable improvement in performance and usually also memory, as can be seen in Table 2. With more aggressive compression schemes, this would likely improve further.\footnote{For the sake of completeness, we would have liked to benchmark ART and HOT too on these compressed datasets, but we were unable to easily adapt the libraries for this purpose. We believe our errors were due to HOPE outputting null characters which interfere with cstring functions in internal use. We do not believe that this is inherently impossible to fix, though.}

\begin{table}[]
\centering
\resizebox{\linewidth}{!}{
\begin{tabular}{l|l|l|l|l}
\textbf{Build, ns/item} & Wiki      & Twitter   & Examiner  & URL         \\
\hline
RSS          & 44       & 41       & 42       & 67        \\
\hline
RSS+HC       & 152      & 163      & 163      & 317      \\
\hline
\hline
\makecell{\textbf{Lookup, ns} \\ \textbf{(LowerBound)}} & Wiki      & Twitter   & Examiner  & URL         \\
\hline
RSS          & 513       & 406       & 482       & 1428        \\
\hline
RSS+HC       & 375 (513) & 292 (406) & 351 (482) & 1095 (1428) \\
\hline
\hline
\textbf{Memory, MB} & Wiki      & Twitter   & Examiner  & URL         \\
\hline
RSS          & 2.6       & 1.4       & 1.6       & 123.0       \\
\hline
RSS+HC       & 22.8      & 3.7       & 6.1       & 278.8       \\
\end{tabular}}
\caption{Results for the RadixStringSpline with and without its add-on Hash Corrector, operating on HOPE-compressed datasets.}
\end{table}

We also wish to note that these comparisons, while accurate to the current state-of-the-art, also take our competitors somewhat out of their element. Both ART and HOT are designed as secondary indexes, storing considerable additional information in leaf nodes which are not used in this primary index / dictionary encoding scenario. Consequently, it is probably possible to streamline both of these data structures for the reduced task at hand. We suspect that doing this would yield a reasonable improvement in memory consumption, but probably not one of the same order as can be provided by a learned model leveraging last-mile search on the underlying data.

%% file: conclusions.tex
\section{Conclusions}
\label{sec:conclusions}

In this work, we have introduced a novel learned string index for the purposes of primary indexing or dictionary encoding which is competitive with existing string index structures in speed and superior in size. We evaluated the learned index on datasets varied in both distribution and size, and found that our method works especially well in conjunction with string compression schemes.

\sparagraph{Future Work.} There exist many interesting future directions: First, better compression techniques could further improve the performance of both RSS and other index structures. Second, the internal redirector of RSS could be improved to be more efficient for large datasets with common prefixes (like URL). Third, RSS currently only uses splines as the main model. However, other types of models could provide significant benefits. In fact, one might see the tree and redirector structure of RSS as a new model for generating error-bounded models out of unbounded components.\footnote{RadixSpline is usually error bounded but is not in this scenario due to the possibility of many duplicate partial keys.} Consequently, this general approach might also be applied to numerical keys to achieve greater space-efficiency. Finally, we also believe there is likely considerable further tuning (and auto-tuning) which could be done on RSS to further improve performance.